\documentclass[conference]{IEEEtran}

\usepackage{graphicx}

\usepackage{cite}
\usepackage{bm,bbm}
\usepackage{algorithm,algpseudocode}
\usepackage{booktabs}
\usepackage{multirow}

\makeatletter
\renewcommand{\ALG@name}{Protocol}
\makeatother

\algnewcommand\algorithmicinput{\textbf{Input:}}
\algnewcommand\Input{\item[\algorithmicinput]}

\algnewcommand\algorithmicinitialization{1: \texttt{Pre-training Process:}}
\algnewcommand\Initialization{\item[\algorithmicinitialization]}

\algnewcommand\algorithmicflprocess{2: \texttt{FL process:}}
\algnewcommand\FLprocess{\item[\algorithmicflprocess]}

\algnewcommand\algorithmicthree{3: \texttt{FL process} is iterated for multiple rounds.}
\algnewcommand\Three{\item[\algorithmicthree]}

\algnewcommand\algorithmicclientupload{4: \texttt{Client Upload:}}
\algnewcommand\ClientUpload{\item[\algorithmicclientupload]}

\algnewcommand\algorithmicaggregationfl{5: \texttt{Aggregation:}}
\algnewcommand\Aggregation{\item[\algorithmicaggregationfl]}

\algnewcommand\algorithmicseven{}
\algnewcommand\Seven{\item[\algorithmicseven]}

\algnewcommand\algorithmicpreini{\texttt{Initialization step:}}
\algnewcommand\PreIni{\item[\algorithmicpreini]}

\algnewcommand\algorithmictraining{\texttt{Training step:}}
\algnewcommand\Training{\item[\algorithmictraining]}

\algnewcommand\algorithmicpruning{\texttt{Pruning step:}}
\algnewcommand\Pruning{\item[\algorithmicpruning]}

\algnewcommand\algorithmicresetting{\texttt{Resetting step:}}
\algnewcommand\Resetting{\item[\algorithmicresetting]}

\usepackage{url}

\usepackage{amsmath,amssymb,amsfonts,mathtools,amsthm}

\usepackage{comment}

\theoremstyle{definition}
\theoremstyle{definition}

\ifCLASSOPTIONcompsoc
	\usepackage[font=normalsize,labelfont=sf,textfont=sf]{subfig}
\else
	\usepackage[font=footnotesize]{subfig}
\fi

\begin{document}
\title{Lottery Hypothesis based Unsupervised Pre-training for Model Compression in Federated Learning}

\author{
\IEEEauthorblockN{
\normalsize Sohei Itahara\IEEEauthorrefmark{1},
\normalsize Takayuki Nishio\IEEEauthorrefmark{1},
\normalsize Masahiro Morikura\IEEEauthorrefmark{1}, and
\normalsize Koji Yamamoto\IEEEauthorrefmark{1}
}
\IEEEauthorblockA{
\IEEEauthorrefmark{1}\small Graduate School of Informatics, Kyoto University,
Yoshida-honmachi, Sakyo-ku, Kyoto 606-8501, Japan\\
Email: nishio@i.kyoto-u.ac.jp
}
}

\maketitle
\begin{abstract}
	Federated learning (FL) enables a neural network (NN) to be trained  using privacy-sensitive data on mobile devices while retaining all the data on their local storages.
	However, FL asks the mobile devices to perform heavy communication and computation tasks, 
	i.e., devices are requested to upload and download large-volume NN models and train them.
	This paper proposes a novel unsupervised pre-training method adapted for FL, which aims to reduce both the communication and computation costs through model compression.
	Since the communication and computation costs are highly dependent on the volume of NN models, reducing the volume without decreasing model performance can reduce these costs.
	The proposed pre-training method leverages unlabeled data, which is expected to be obtained from the Internet or data repository much more easily than labeled data.
	The key idea of the proposed method is to obtain a ``good'' subnetwork from the original NN using the unlabeled data based on the lottery hypothesis.
	The proposed method trains an original model using a denoising auto encoder with the unlabeled data and then prunes small-magnitude parameters of the original model to generate a small but good subnetwork.
	The proposed method is evaluated using an image classification task.
	The results show that the proposed method requires 35\% less traffic and computation time than previous methods when achieving a certain test accuracy.
\end{abstract}
\IEEEpeerreviewmaketitle

\section{Introduction}
Deep learning has dramatically improved the performance of many AI products such as image recognition, language processing, and voice recognition.
To obtain such high-performance models, huge amounts of data and computation are required for model training.
A new trend has arisen to utilize the data and computation power on networks, i.e., resources on mobile devices such as smartphones, laptops, and connected vehicles that are equipped with powerful sensors and computation resources.

Federated learning (FL)~\cite{FedAve} is a novel learning mechanism used to train models by utilizing the data and computation power of the devices without collecting their data, which is often privacy-sensitive, in a central training system.
Google has leveraged FL to train a neural network (NN) model for Google keyboard query suggestions with mobile users' data~\cite{Google_FL}.
The FL process has several specific properties as follows:
1) Limited bandwidth: mobile devices are often connected to networks through speed limited wireless links.
2) Limited computation capacity: 
mobile devices are equipped with weak computation resources, which are often used for the other applications simultaneously.
As a result, only limited computation capacity is available for the FL process.
3) Non-independent and identically distributed (non-IID) data: the local dataset on each device typically depends on the usage of the device by its user, and hence the user's local dataset will not be representative of the population distribution.
We represent the nature of data distribution as non-IID.

\begin{figure}[!t]
    \centering
    \includegraphics[width = 0.4\textwidth]{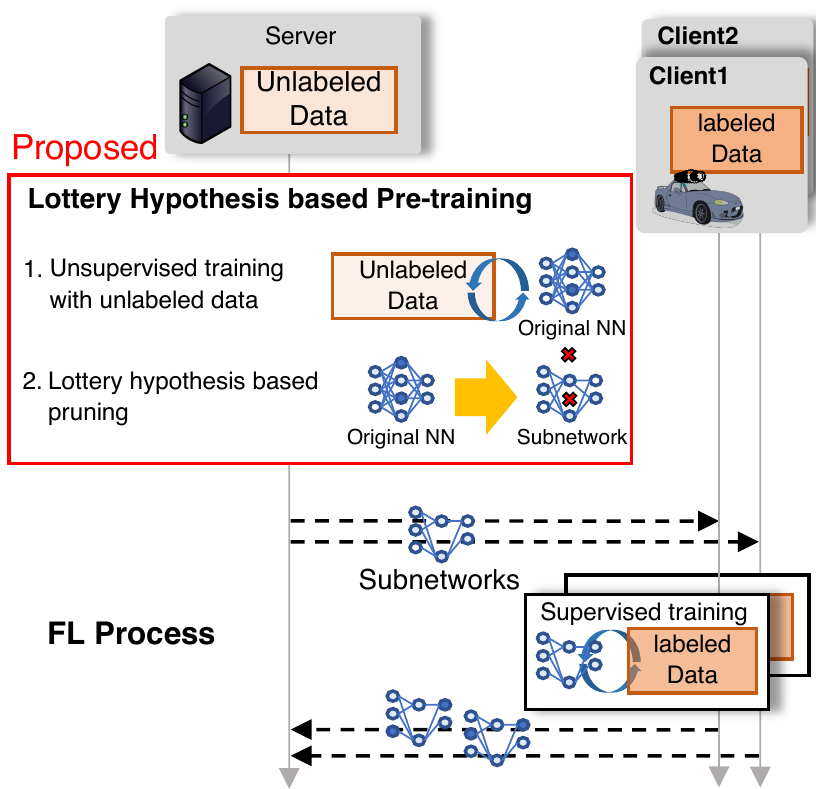}
	\caption{\textbf{Overview of the proposed HP-FL.}
	 The dotted black lines indicate communications with the server and mobile devices.
	 The solid black squares indicate the computation on the mobile devices.
	 The solid red square indicates the proposed pre-training method.
	 }
    \label{fig:HP-FL}
\end{figure}

In FL, each device updates a NN model with its data and uploads the model to the server instead of uploading the data. 
Thus, heavy computation and traffic are required when the model requested to train is huge, i.e., the model has a high number of parameters.
The heavy computation and communication could discourage mobile device users from participating in the FL.

This paper proposes a novel pre-training method adapted for FL, named hybrid pre-training federated learning (HP-FL), which aims to reduce the communication and computation costs of the devices in the FL process.
Since these costs are highly dependent on the volume of NN models, they can be reduced by reducing the volume without decreasing model performance. 
The key idea of the proposed method is to obtain a small but ``good'' subnetwork from the original NN using the unlabeled data based on the lottery hypothesis~\cite{lottery_hypothesis}.
In this method, in addition to the local data of the mobile devices, unlabeled open data is used for the pre-training process on the server, which are obtained from the Internet, specifically, in the web pages or data repositories.
The HP-FL pre-trains and compresses an original NN model with the unlabeled data so that the communication and computation costs in the clients are reduced.

The contributions of this paper are summarized as follows:

\begin{itemize}
	\item 
	We propose a novel unsupervised pre-training method adapted for FL.
	To the best of our knowledge, this is the first work discussing the pre-training method adapted for FL.
	It uses unsupervised training of a denoising auto encoder (DAE) and parameter pruning based on the lottery hypothesis to obtain a compressed model using only unlabeled data without sacrificing the accuracy.

	\item 
	We evaluate our method using an object classification task MNIST.
	The results show that our method realized 35\% lower communication and computation costs than existing methods, with only a slight reduction in accuracy (1.3\%).
\end{itemize}

\section{Related Works}

\subsubsection{Model Compression in FL}

Model compression is a promising method to reduce communication traffic and computation cost.
There are two approaches to model compression in FL: quantization and sparsification. 
For the centralized training, Li et al. proposed a method to train ternary weight networks~\cite{TWN}, which quantized the parameters of a NN to three values.
For the decentralized training, Bernstein et al. proposed signSGD~\cite{signsgd}, which quantized every gradient upload to its binary sign, thus reducing the bit size per upload by a factor of $32$.
Quantization can reduce the size of the data in the trained model, but the computation cost is not always reduced.

For sparsification, Aji and Heafield proposed a method~\cite{sparseFL}, where the server fixes the sparsity rate $p$ and the clients only upload the fraction $p$ entries with the highest magnitude of each gradient. 
Additionally, Sattler et al. proposed sparse ternary compression~\cite{STC_FL}, which integrates quantization and sparsification using an effective coding method.
In these methods, the original model, which is before compressed, is used in the model update on the client, and then, the clients compress the model and upload it. 
This results in a reduction in the communication cost, however, the computation cost is not reduced.
On the other hand, our proposed method reduces not only the communication cost but also the computation cost.

\subsubsection{Denoising Auto Encoder}

A DAE~\cite{DAE} is a kind of auto encoder that can learns representations that are robust to partial corruption of the input pattern and is used for unsupervised training to initialize a NN.
Consider a NN $\bm{y} = f(\bm{x};\bm{\theta})$,
 which consists of $N$ layers and is parametrized $\bm{\theta} = \{\bm{W}_i,\bm{b}_i\}_{i=0}^N$,
 where $\bm{W}_i$ is a $d_i \times d_{i+1}$ weight matrix and $\bm{b}_i$ is a bias vector of the $i$-th layer. 
Consider the initialization of the NN using a DAE.
The DAE contains an encoder $\bm{y} = f(\bm{x};\bm{\theta})$ and a decoder $\bm{z} = g(\bm{y};\bm{\theta'})$;
 the encoder has a same architecture as the NN which will be initialized,
 and the decoder consists of $N$ layers and is parametrized $\bm{\theta} = \{\bm{W}_i',\bm{b}_i'\}_{i=0}^N$,
 where $\bm{W}_i'$ is a $d_{N+2-i} \times d_{N+1-i}$ weight matrix and $\bm{b}_i'$ is a bias vector of the $i$-th layer. 
Let us denote the DAE as $\bm{z} = h(\bm{x};\bm{\Theta}) = g(f(\bm{x};\bm{\theta});\bm{\theta}')$,
 where $\bm{\Theta} = \{\bm{\theta},\bm{\theta}'\}$.

The previous method to train DAEs is a layer-wise procedure~\cite{stack_DAE}, which avoids gradient disappearance.
The DAE is trained to reconstruct a clean repaired input from a corrupted one, using only unlabeled data.
First, the input $\bm{x}$ is corrupted, which results in $\hat{\bm{x}}$.
The parameters of a DAE $\bm{\Theta}$ are updated as follow:
\begin{eqnarray}
	\bm{\Theta} \leftarrow \bm{\Theta} - \eta \nabla L(h(\hat{\bm{x}};\bm{\Theta}),\bm{x}),
\end{eqnarray} 
where $L$ is a loss function and $\eta$ is a learning rate.
After this unsupervised training, weight matrix and bias vector of each layer of the encoder is used as initial values of corresponding layer for the NN, which is trained in supervised learning process.

\subsubsection{Lottery Hypothesis}
The lottery hypothesis~\cite{lottery_hypothesis} states that a randomly-initialized NN contains a small but ``good'' subnetwork that is initialized such that, when trained in isolation, it can match the test accuracy of the original network after training.
To explain in more detail, consider an original NN $f(\bm{x};\bm{\theta})$ and a mask $\bm{m} \in {\{0,1\}}^{|\bm{\theta}|}$,
 let a subnetwork defined by the original network $f(\bm{x};\bm{\theta})$ and the mask $\bm{m}$ be denoted as $f(\bm{x};\bm{m}\odot\bm{\theta})$,
 where $\bm{m}\odot\bm{\theta}$ is element-wise product.
In the training process of the subnetwork, only the masked parameters are updated.
More formally, when a certain element of the mask is 1, the corresponding parameter is updated.
Conversely, when a certain element is 0, the corresponding parameter is fixed to 0.
Through such training process, when we select an appropriate mask corresponding to the initial value $\bm{\theta}_0$, the subnetwork $f(\bm{x};\bm{m}\odot\bm{\theta}_0)$ achieves competitive accuracy to the original network achieves.

\section{Proposed Method: Hybrid Pre-training (HP-FL)}
\label{sec:system_model}
\subsection{Assumption}
The system model consists of a server and many clients (e.g., millions of mobile devices such as smartphones, tablets, and connected cars), which are connected through wireless or wired networks. 
We assume these clients are equipped with a considerable weaker computation capacity than the server.  

As with other FL research, in this study, we assume clients have a small amount of labeled data.
The clients' data are typically generated and labeled by the usage of the mobile device; hence their data are non-IID.
Additionally, we assume the server holds the unlabeled open data.
On the Internet, specifically in the publicly available data repositories, there is a huge amount of input data.
Although it is unlikely that the data on the repository will have desirable labels, the unlabeled data can be collected from the Internet much more easily than labeled data.
The labeling cost is high in a supervised learning process where the labeling is performed by a human,
 for example, support to train a classifier to discriminate between attractive and unattractive photos using users' privacy-sensitive photos on devices. 
Devices such as smartphones hold photos and corresponding attractiveness labels (e.g., how often a photo is displayed or shared).
On the other hand, the server obtains many open photos on a web site (e.g., Google Image), however, they do not have attractiveness labels.

\subsection{HP-FL protocol}

We propose a novel pre-training method adapted for FL, named \texttt{HP-FL}.
As shown in Protocol~\ref{protocol:HP-FL}, \texttt{HP-FL} contains two processes, which are performed sequentially: \texttt{pre-training process} and \texttt{FL process}.
In the \texttt{pre-training process}, a small but ``good'' subnetwork of the original NN is extracted using unlabeled data on the server.
In the following \texttt{FL process}, the clients download, update, and upload only the subnetwork, which results in a reduction in the communication and computation costs simultaneously.

\begin{algorithm}[h]
    \caption{Hybrid pre-training federated learning (\texttt{HP-FL}).}         
    \label{protocol:HP-FL}
	\begin{algorithmic}[]

    \Initialization
	\State A server obtains a small but ``good'' subnetwork $\bm{m}\odot\bm{\theta}_0$ (Protocol~\ref{protocol:pre-training} describes this in more detail).
	\\
	\FLprocess
	\State \texttt{Download step}
	\State Each client downloads the parameters of subnetwork.
	\State \texttt{Update step}
	\State Each client updates the subnetwork using its data.
	\State \texttt{Upload step}
	\State Each client uploads the parameters of updated subnetwork to the server.
	\State \texttt{Aggregation step}
	\State The server averages the uploaded parameters to renew the subnetwork.
    \\
    \Three 
    \end{algorithmic}
\end{algorithm}

Consider an original NN $f(\bm{x};\bm{\theta})$, which will be compressed to a small subnetwork. 
The DAE is represented as $h(\bm{x};\bm{\Theta}) = g(f(\bm{x};\bm{\theta});\bm{\theta}')$,
 where $\bm{y} = f(\bm{x};\bm{\theta})$ is the same network architecture as the original NN model,
 which is called as an encoder, and $g(\bm{y};\bm{\theta}')$ is called as a decoder.
The DAE is trained with unlabeled dataset and the parameters of the encoder network $\bm{\theta}$ is used as an initial parameter of the original NN model.
Then, the model is pruned to obtain a small subnetwork used in \texttt{FL process}.

The details of the proposed \texttt{pre-training process} are shown in Protocol~\ref{protocol:pre-training}.
First, in the \texttt{Initialization step}, we randomly initialize the parameters of the DAE $\bm{\Theta}_0 = (\bm{\theta}_0,\bm{\theta}_0')$ and set all the elements of a DAE mask $\bm{M} = (\bm{m},\bm{m}')$ to 1.
The DAE mask comprises two parts: an encoder mask $\bm{m} \in \{0,1\}^{|\bm{\theta}|} $ and a decoder mask $\bm{m}' \in \{0,1\}^{|\bm{\theta'}|}$.
The DAE mask $\bm{M}$ determines the trainable parameters of the DAE, which is updated in the following \texttt{Training step}.
More formally, when a certain element of the mask is 1, the corresponding parameter is updated.
Conversely, when a certain element is 0, the corresponding parameter is fixed to 0.
The DAE mask $\bm{M}$ determines the subnetwork of the DAE $h(\bm{x};\bm{M}\odot\bm{\Theta})$, where $\bm{M}\odot\bm{\Theta} \coloneqq (\bm{m}\odot\bm{\theta},\bm{m}'\odot \bm{\theta}') $.

\begin{algorithm}[h]
    \caption{\texttt{Pre-training process}.}         
    \label{protocol:pre-training}         
	\begin{algorithmic}[]
	\Input{DAE: $h(\bm{x};\bm{\Theta})$,
	DAE masks: $\bm{M}$, unlabeled data: $\bm{x}$, the number of iterations: $N_\mathrm{p}$, pruning rate: $p$\%}
	\\
    \PreIni
	\State Randomly initialize $\bm{\Theta}_0$ and initialize $\bm{M}\in \{1\}^{|\bm{\Theta}|}$
	\Training
	\State Add noise $\bm{\epsilon}$ to $\bm{x}$, which results in $\bm{\hat{x}}$.
	\State $\bm{\Theta} \leftarrow \bm{\Theta} - \eta \nabla L(h(\hat{\bm{x}};\bm{M}\odot\bm{\Theta}_0),\bm{x})$
	\Pruning
	\State Prune $p$\% of the smallest magnitude of survived parameters $\bm{M}\odot\bm{\Theta}$.
	\Resetting
	\State Reset the remaining parameters to their values in $\bm{\Theta}_0$.
	\\
	\Seven{All steps except the \texttt{Initialization} step are iterated for $N_\mathrm{p}$ iteration.}
    \end{algorithmic}
\end{algorithm}

In the \texttt{Training step}, we update the subnetwork of DAE using unlabeled data on the server.
The subnetwork of the DAE is trained to reconstruct a clean input $\bm{x}$, from a corrupted input $\hat{\bm{x}}$.
Previous methods to train the DAE include a layer-wise procedure~\cite{stack_DAE}, to avoid gradient disappearance.
However, the entire network is trained as with usual NN training in this paper because the NN used in this paper is not so deep and can be trained well by the end-to-end procedure.

In the \texttt{Pruning step}, $p$\% of parameters in the original NN are pruned. 
In other words, we select $p$\% of the smallest magnitude of the parameters of the subnetwork and substitute 0 for all the elements of the DAE mask that correspond to selected parameters.
In the \texttt{Resetting step}, the remaining parameters are reset to their values in $\bm{\Theta}_0$.
A straightforward approach to prune parameter is one-shot: the DAE is trained once, some weights are pruned, and the surviving weights are reset.
However, we conduct iterative pruning, which repeatedly trains, prunes, and resets the network over ${N_\mathrm{p}}$ iterations; each iteration prunes $p$\% of the weights that survive from the previous pruning step.
The lottery hypothesis~\cite{lottery_hypothesis} states that iterative pruning finds a subnetwork that matches the accuracy of the original network at smaller sizes than one-shot pruning.

After this iterative pruning, we obtain the masks $\bm{M}=(\bm{m},\bm{m}')$ and the initial weights $\bm{\Theta}_0 = (\bm{\theta}_0,\bm{\theta}_0')$ of the DAE.
In the following \texttt{FL process}, we use the subnetwork $f(\bm{x},\bm{m}\odot\bm{\theta}_0)$, which is determined by the encoder mask $\bm{m}$ and the initial value of encoder $\bm{\theta}_0$.
Additionally, we denote the ratio of survived parameters $|\bm{m}|^{0}/|\bm{m}|$ as the remaining rate $P_\mathrm{R}$, where $|\bm{m}|^0$ represents the number of parameters of the subnetwork and $|\bm{m}|$ represents the number of parameters of the original NN.

In the following \texttt{FL process}, the clients download, update, and upload only the subnetwork, which is made in the \texttt{pre-training process}.
In more detail, the clients download, update, and upload only the parameter, whose corresponding elements of the mask is 1.  
Comparing the communication and computations cost per round required for \texttt{HP-FL} and FL, which uses the original network,
 the costs of \texttt{HP-FL} was reduced to $P_\mathrm{R}$ times that of FL.

\section{Performance Evaluation}
\label{sec:evaluate}
\subsection{Evaluation Setup}

\subsubsection{Data Distribution}

In this evaluation, we used MNIST~\cite{lenet}, which is a major image classification task.
MNIST is a classic object classification dataset, consisting of 60,000 training images and 10,000 testing images with 10 image classes. 

The data distribution over the clients was non-IID and determined as follows.
First, we fixed the number of clients $K$ to 100.
Then the data were divided into the unlabeled open data consisting of 20,000 images, which used in the \texttt{pre-training process}, and the labeled data consisting of 40,000 image--label pairs, which used in \texttt{FL process}.
To distribute the labeled data to the clients, the data was sorted by its classification label, divided into 200 shards of size 200, and then each client was assigned two shards.
With this data distribution, most clients will only have the 400 data samples consisting of two classes.

\subsubsection{ML Model and its Hyperparameter}

In the \texttt{FL process}, we used Lenet~\cite{lenet}, which consists of three fully connected layers (300, 100 units with ReLU activation, and 10 units activated by soft-max activation), resulting in 266,610 parameters (1.1 MB in 32-bit float).
When updating the model in the \texttt{Update step} of the \texttt{FL process}, we selected hyperparameters as follows: 60 for mini-batch size, 5 for the number of epochs in each round, 0.1 for the learning rate, stochastic gradient descent (SGD) for the optimizer, and categorical cross entropy for the loss function.
We assume the time required to update the original neural network (Lenet) on each client is $t_\mathrm{comp} = 10\,\mathrm{s}$.
We considered this to be reasonable because our workstation (GeForce GTX 1080 Ti) required 0.1\,s for a single update with a single GPU;
 mobile devices with a weaker computation resource could require an update time 100 times longer.
The result is that when the client updates the subnetwork with the remaining rate $P_\mathrm{R}$, the time required to update is $10P_\mathrm{R}\,(\mathrm{s})$.

The DAE used in the \texttt{pre-training process} consists of an encoder and a decoder. 
The encoder has the same architecture as Lenet, consisting of three fully connected layers (300, 100 units with ReLU activation, and 10 units activated by linear activation). 
The decoder consists of three fully connected layers (100, 300 units with ReLU activation, and 784 units activated with sigmoid activation).
To train the DAE, we make corrupted unlabeled data $\hat{\bm{x}}$ from clean unlabeled data $\bm{x}$ as follows.
First, we add Gaussian noise $\epsilon \sim \mathcal{N}(0.5,0.25)$ to each element of $\bm{x}$ and clip them into the range 0--1.
When updating the DAE in the \texttt{Training step}, we selected hyperparameters as follow:
 100 for mini-batch size, 100 for the number of epochs in each \texttt{Training step}, 0.001 for the learning rate, adam for the optimizer, and mean squared error for the loss function.
In the \texttt{Pruning} step, we select 20\% for the pruning rate $p$. 

\subsubsection{Evaluation Details}

We compared \texttt{HP-FL} with the following FL protocols:
\texttt{Original FL} without pre-training and pruning, which uses a randomly initialized original network for the \texttt{FL process}.
\texttt{FL} with pre-training, which uses the pre-trained original network, 
 i.e., we perform only the \texttt{Initialization step} and \texttt{Training step} for a single iteration in \texttt{pre-training process} and then use trained parameters of encoder $\bm{\theta}$ as the pre-trained initial value.
\texttt{FL} with random pruning, which uses a randomly selected subnetwork.
All the methods use the same architecture (Lenet) and the same initial value for the \texttt{pre-training process}.

\texttt{HP-FL} and the other three methods are evaluated based on the following metrics.
\begin{itemize}
	\item \textbf{Communication cost or computation time required to achieve a desired accuracy} (CCA@$x$, CTA@$x$):
	we evaluate the communication cost of the server and computation time on each client to achieve a certain accuracy $x$ (i.e., the lower the better).
	We denote the number of rounds of the \texttt{FL process} required to achieve the accuracy $x$ as NRA@$x$.
	The communication cost to achieve an accuracy $x$ (CCA@$x$) is numerically calculated as $|\bm{\theta}| \times P_\mathrm{R} \times 2K \times$ NRA@$x$.
	The computation time to achieve an accuracy $x$ (CTA@$x$) is numerically calculated as $t_\mathrm{comp} \times P_\mathrm{R} \times$NRA@$x$.
	\item \textbf{Accuracy after enough rounds} (\textit{Accuracy}):
	 we also measured the accuracy on testing datasets after 3000 rounds of the \texttt{FL process}.
\end{itemize}

\subsection{Evaluation Results}

\begin{table*}[t]
	\caption{\textbf{Results of \texttt{HP-FL} and the other three methods.}
	NRA@$x$: the number of rounds required to achieve a testing classification accuracy of $x$.
	CCA@$x$, CTA@$x$: the communication cost or computation time  required to achieve an accuracy of $x$ (i.e., the lower the better).
	\textit{Accuracy}: the testing accuracy after 3000 rounds (i.e., the higher the better).
	For both \texttt{HP-FL} and \texttt{FL} with random pruning, the remaining rate $P_\mathrm{R}$ is 0.107, i.e., the size of subnetwork is about 10 times smaller than the original network
	\texttt{FL} with pre-training and \texttt{Original FL} without pre-training and pruning use the original model.
	}
    \label{table:main}
    \centering
    \begin{tabular}{cccccccc}
		\toprule
		\multirow{2}{*}{Method}& Communication&Computation& \multirow{2}{*}{NRA@95\%} &\multirow{2}{*}{CCA@95\%}& \multirow{2}{*}{CTA@95\%} & \multirow{2}{*}{\textit{Accuracy}} \\
		&Cost / Round&Time / Round\\
		\midrule
		\textbf{Proposed: \texttt{HP-FL}} with pre-training and pruning&22.8\,MB&1.1\,s&824&\textbf{18.8\,GB}&\textbf{14.7\,min}&\textbf{96.3\%}\\
		\texttt{FL} with random pruning &22.8\,MB&1.1\,s&1388&31.7\,GB&24.8\,min&95.7\%\\
		\texttt{Original FL} without pre-training and pruning &213\,MB&10\,s&134&29.2\,GB&22.5\,min&97.6\%\\
		\texttt{FL} with pre-training &213\,MB&10\,s&126&29.0\,GB&22.3 min&97.2\%\\
        \bottomrule
    \end{tabular}
\end{table*}

\begin{table*}[t]
	\caption{\textbf{Effect of different remaining rates $P_\mathrm{R}$}.
	Here, $P_\mathrm{R}$ determines the size of the subnetwork that is downloaded, uploaded, and updated by the clients in the \texttt{FL Process}.}
    \label{table:P_R_effect}
    \centering
    \begin{tabular}{cccccccc}
		\toprule
		\multirow{2}{*}{$P_\mathrm{R}$} &Size of& Communication&Computation& \multirow{2}{*}{NRA@95\%}&\multirow{2}{*}{CCA@95\%}& \multirow{2}{*}{CTA@95\%}&\multirow{2}{*}{\textit{Accuracy}}\\
		&subnetwork&Cost / Round&Time / Round\\
		\midrule
		1.0 & 267k &  213\,MB & 10\,s  & 134 & 29.2\,GB & 22.3\,min & 97.6\%\\
		0.41 & 109k&  87.3\,MB& 4.1\,s & 267 & 23.4\,GB & 18.2\,min & 97.3\%\\
		0.21 & 56k &  44.7\,MB& 2.1\,s & 513 & 23.0\,GB & 18.0\,min & 96.7\%\\
		0.107 & 29k & 22.8\,MB & 1.1\,s & 824 & 18.8\,GB & 14.7\,min & 96.3\% \\
        \bottomrule
    \end{tabular}
\end{table*}

The main results are listed in Table~\ref{table:main}.
\texttt{HP-FL} outperforms the other methods in terms of CCA and CTA, with a slight reduction in \textit{Accuracy}.
One reason for the improvement of CCA and CTA is that the communication cost and computation cost per round in \texttt{HP-FL} is much lower than that of the two methods that use the original network, as shown in Table~\ref{table:main} as Communication Cost / Round and Computation Time / Round.
This is because, in \texttt{HP-FL}, the clients communicate and compute only the subnetwork, which is compressed from the original network in the \texttt{pre-training process}, whereas in the other two methods, the clients use the original network.

On the other hand, to compare \texttt{HP-FL} with \texttt{FL} with random pruning, \texttt{HP-FL} performs the best in terms of CCA, CTA, and \textit{Accuracy},
 while the communication and computation costs per round are the same.
This is because to train the randomly pruned network requires more rounds than the network pruned by the proposed method, which is shown in Table~\ref{table:main} as NRA@95\%.
This means that \texttt{HP-FL} obtains a ``better'' subnetwork in the \texttt{pre-training process} than at least randomly compression.

Table~\ref{table:P_R_effect} lists the effect of the remaining rate $P_\mathrm{R}$, which determines the number of parameters of the subnetwork used in the \texttt{FL process} (the lower $P_\mathrm{R}$, the smaller subnetwork).
In the case of $P_\mathrm{R}=0.41$, \texttt{HP-FL} achieves 20\% less CCA and CTA, without sacrificing the \textit{Accuracy}.
The results listed in Table~\ref{table:P_R_effect} demonstrate that a lower $P_\mathrm{R}$ results in lower CCA@95\%, CTA@95\% (the lower the better), and lower \textit{Accuracy} (the higher the better).
When the size of the subnetwork is $P_\mathrm{R}$ times smaller than the original network, 
 CCA and CTA are larger than $P_\mathrm{R}$ times that required by \texttt{Original FL}.  
This is because the smaller the subnetwork, the more rounds required to obtain a certain accuracy, as shown in Table II as NRA@95\%.

\section{Conclusion}
\label{sec:conclusion}
We have presented a new pre-training method, \texttt{HP-FL}, which aims to reduce both the communication and computation cost of FL.
The key idea is to compress the NN based on the lottery hypothesis before the FL process using only unlabeled data.
Our evaluation results have revealed that \texttt{HP-FL} provided a high-performance and compressed NN, while requiring a significantly lower communication and computation costs compared with the previous methods.
An interesting area for future work will be to compress the NN, leveraging not only unlabeled data on the server but also labeled privacy-sensitive data on the mobile devices.

\section*{Acknowledgment}
This work was supported in part by JSPS KAKENHI Grant Number JP18K13757 and KDDI Foundation.

\bibliographystyle{IEEEtran}
\bibliography{vtcf_2020}
\end{document}